# On the isotope effect in compressed superconducting H$_3$S and D$_3$S

**Dale R. Harshman[1] and Anthony T. Fiory[2]**

[1] Department of Physics, The College of William and Mary, Williamsburg, Virginia 23185, USA

[2] Department of Physics, New Jersey Institute of Technology, Newark, New Jersey 07102, USA

E-mail: drh@physikon.net

**Abstract**

A maximum superconductive transition temperature $T_C$ = 203.5 K has recently been reported for a sample of the binary compound tri-hydrogen sulfide (H$_3$S) prepared at high pressure and with room temperature annealing. Measurements of $T_C$ for H$_3$S and its deuterium counterpart D$_3$S have suggested a mass isotope effect exponent α with anomalous enhancements for reduced applied pressures. While widely cited for evidence of phonon-based superconductivity, the measured $T_C$ is shown to exhibit important dependences on the quality and character of the H$_3$S and D$_3$S materials under study; examination of resistance versus temperature data shows that variations in $T_C$ and apparent α are strongly correlated with residual resistance ratio, indicative of sensitivity to metallic order. Correlations also extend to the fractional widths of the superconducting transitions. Using resistance data to quantify and compensate for the evident materials differences between H$_3$S and D$_3$S samples, a value of α = 0.043 ± 0.140 is obtained. Thus, when corrected for the varying levels of disorder, the experimental upper limit (≤0.183) lies well below α derived in phonon-based theories.

## Introduction

The highest reported superconducting transition temperature is presently found in highly compressed tri-hydrogen sulfide [1] (H$_3$S, symmetry group Im$\bar{3}$m [2]), exhibiting a diamagnetism onset $T_C$ ~ 200 K at applied hydrostatic pressure $P$ = 155 GPa, with comparative data on D$_3$S showing $T_C$ ~ 150 K [1]. Taken at face value, these data suggest a finite H-D-mass isotope effect with exponent α ~ 0.4 and a phonon-based origin for the superconductivity [1].

Prior to its discovery, calculations based on strong electron-phonon coupling and Eliashberg theory predicted $T_C$ ~ 200 K for H$_3$S [3]. Post-discovery formalisms, which consider both harmonic and anharmonic effects in the H vibrations, produce similar findings for H$_3$S, derive $T_C$ ~ 150 K for D$_3$S, and calculate α = 0.35 or 0.38 ≤ α ≤ 0.42 [4-7]. Comparisons made with experiment, such as an apparent divergence in α at low $P$ [8], generally presume that the differences between H$_3$S and D$_3$S are intrinsic and free from materials effects. As is shown below, however, interpreting the experimental data is most certainly complicated by the evident effects of disorder on $T_C$, which are comparatively more prevalent in D$_3$S samples.

Of particular interest to the present study are data for samples of H$_3$S and D$_3$S annealed at temperature $T$ at or above room temperature and $P \gtrsim$ 150 GPa; this process is found to reduce the normal-state resistance, while increasing $T_C$ [1,2]. In H$_3$S, thermal annealing induces a kinetically controlled phase transformation, as deduced from observations of sharp increases of $T_C \gtrsim$ 70 K for $P \gtrsim$ 160 GPa [1]. However, such annealing evidently does not completely remove the influence of materials disorder in the samples. The highest $T_C$ = 203.5 K reported for H$_3$S at $P$ = 155 GPa was determined by temperature extrapolation from magnetic field sweep measurements. For the same sample, temperature dependence in zero-field-cooled magnetization and electrical resistivity indicate lower superconducting onset temperatures, 200.4 K and 195.8 K, respectively, while zero resistance appears to set in below 170 K (as read from figure 4(a) of [1]). Field-cooled magnetization data exhibit no evidence of the superconducting transition or Meissner flux exclusion, which has been attributed to formation of fluxons strongly pinned to material defects.

## Methods and results

Obtained from measurements of resistance and magnetic transitions, results for $T_C$ indicate a maximum in the dependence on $P$, along with an H-D isotope dependent difference, as shown in figure 1. Data for $T_C$ versus $P$ are read from figure 2(c) of [1] and figure 3(c) of [2]. Additional data for figure 1 are $T_C$ determined from superconducting onset temperatures in resistance versus temperature $R(T)$ traces shown in figures 2(b), 3(b), and 4(a) of [1] and figures 3(a) and 3(b) of [2], together with the associated values of $P$. As indicated in figure 3(c) of [2], the regions of highest $T_C$ appear to distinguish R3m crystal structure at lower $P$ and Im$\bar{3}$m structure at higher $P$, although a structural distinction was not clearly discernible from these x-ray diffraction experiments.

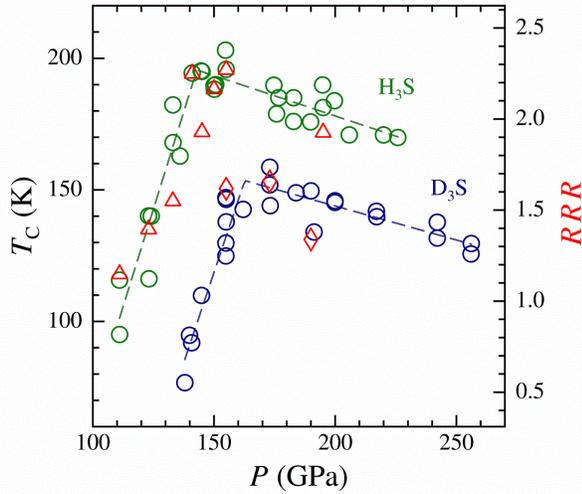

**Figure 1.** Transition temperatures $T_C$ for $H_3S$ (green circles) and $D_3S$ (blue circles) versus applied pressure $P$ with corresponding dashed lines fitted to joined linear trends using equation (1) with parameters in table 1 (left scale). Residual resistance ratios $RRR$ are shown for $H_3S$ (red triangles) and $D_3S$ (red diamonds) versus $P$ (right scale). Data are for samples annealed at or above room temperature [1,2].

Despite substantial scatter among the data points in figure 1, several qualitative features are clearly evident: (1) $T_C$ for $H_3S$ is generally higher than $T_C$ for $D_3S$; (2) the region of maximum $T_C$ for $H_3S$ occurs at lower $P$ than for $D_3S$; and (3) $T_C$ shows pronounced linear-like variations at low $P$ and comparatively shallow linear-like variations at high $P$. These characteristics can be represented by modeling $T_C$ as a piecewise linear function of $P$,

$$T_C(p) = T_m [1 - a(1-p)\,\theta_{hv}(1-p) - b(p-1)\,\theta_{hv}(p-1)], \quad (1)$$

written in terms of reduced pressure $p = P/P_m$, corresponding to a maximum temperature $T_m$ at pressure $P_m$, and slopes $a$ and $b$; $\theta_{hv}$ is the Heaviside unit step function. Fits of equation (1) are displayed as the dashed lines in figure 1; the fitted parameters are given in table 1; $\sigma_{fit}$ is the root-mean-square deviation between the function of equation (1) and measured $T_C$ and indicates the experimental reproducibility is 7–8 K. The fitted pressure $P_m$ at the $T_m$-point for $D_3S$ is higher, relative to $H_3S$, by $19.8 \pm 2.6$ GPa, which compares well to the 18 GPa pressure difference associated with the highest $T_C$ measured for each isotope in figure 1.

Information contained in recordings of resistance versus temperature $R(T)$ for samples annealed at room temperature, of which eight are available for $H_3S$ and three for $D_3S$ [1,2], illustrate a correlation between normal-state temperature dependence and the width of the superconducting transition. The temperature dependence is represented by the residual resistance ratio $RRR = R(T_{C0})/R(0)$, using as reference temperatures the highest measured $T_{C0} = 203.5$ K and $T = 0$, as determined from linear extrapolations of $R(T)$ in the normal-state region $T > T_C$, where $R(T)$ is nearly linear in $T$. As customarily applied to metals with room temperature as the upper reference point, $RRR$ provides a specific measure for distinguishing relative order or purity among samples; e.g., optimal high-$T_C$ superconductors in the clean limit generally have $RRR \gg 1$, and depression in $T_C$ has been found to follow depression in $RRR$ [9]. The data for $RRR$ are shown in figure 1, scaled on the right ordinate axis. As is clearly observable, the maxima in the $P$ dependence of $T_C$ are clearly correspondingly correlated with maxima in $RRR$. By way of comparison, $H_3S$ samples not annealed at room temperature display smaller $RRR \lesssim 1.06$ and are semiconductor-like for $P < P_m$ (from figure 1(a) of [1]).

**Table 1.** Values of parameters $T_m$, $P_m$, $a$, and $b$ from fitting equation (1) to data for $T_C$ versus $P$ in figure 1 for $H_3S$ and $D_3S$. Statistical uncertainties in the two least significant digits are given in parentheses; fitting rms deviation is given by $\sigma_{fit}$.

|  | $T_m$ (K) | $P_m$ (GPa) | $a$ | $b$ | $\sigma_{fit}$ (K) |
|---|---|---|---|---|---|
| $H_3S$ | 195.6(3.0) | 142.9(1.9) | 2.16(18) | 0.224(55) | 8.3 |
| $D_3S$ | 153.5(3.0) | 162.7(1.8) | 2.91(26) | 0.274(61) | 6.6 |

The relative transition width $\Delta T_C/T_C$ provides a quantitative measure to compare homogeneity in the superconducting states of various samples. The width of the superconducting transition can be determined from the transition midpoint where $R(T_{½}) = ½\, R(T_C)$ and defined as $\Delta T_C = T_C - T_{½}$. Results from the eleven $R(T)$ measurements are shown as $RRR$ versus $\Delta T_C/T_C$ in figure 2, where the dashed line is a linear fit indicating the trend among the data points. The systematic decreases in $RRR$, accompanied by systematic increases in $\Delta T_C/T_C$, show that the three $D_3S$ samples have lesser metallic order and superconducting homogeneity, when compared to the five $H_3S$ samples with highest order and sharpest superconducting transitions. Applied pressures for these five $H_3S$ samples and the three $D_3S$ samples correspond to $P \gtrsim P_m$. The data for the three $H_3S$ samples with relatively broad superconducting transitions ($\Delta T_C/T_C > 0.05$) were obtained at pressures $P < P_m$.

A corresponding correlation also occurs between $T_C$ and $RRR$, as shown in figure 3. Data available at $P \gtrsim P_m$ for both isotopes are indicated by the eight dot-filled symbols and show the trend given by the fitted diagonal line. The data in both figures 2 and 3 indicate lesser quality among the $D_3S$ samples, which undoubtedly contributes to their lower values of observed $T_C$, relative to $H_3S$.

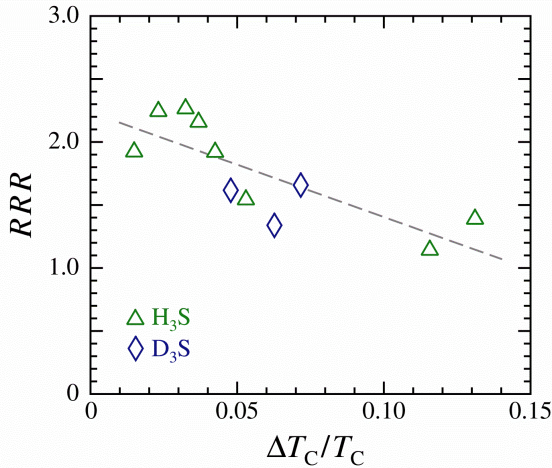

**Figure 2.** Residual resistance ratio $RRR$ versus fractional transition width $\Delta T_C/T_C$ for samples of $H_3S$ (green triangles) and $D_3S$ (blue diamonds). A linearly fitted dashed line is drawn to indicate the trend.

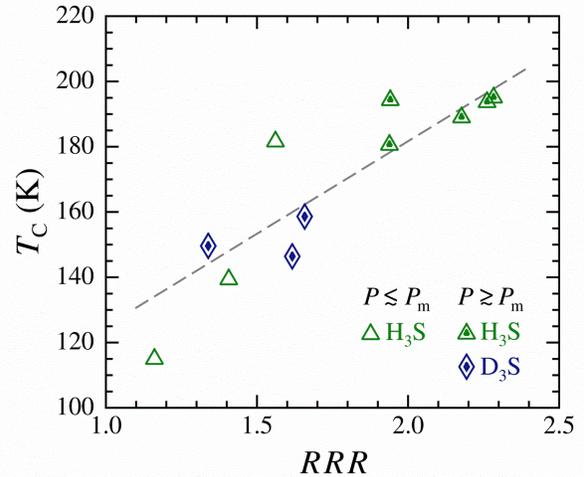

**Figure 3.** Transition temperature $T_C$ versus residual resistance ratio $RRR$ for $H_3S$ (green triangles) and $D_3S$ (blue diamonds). The eight symbols with dot fill correspond to data for $P \gtrsim P_m$ with trend shown by the fitted dashed line.

**Table 2.** Data for applied pressure $P$ and $T_C$ for samples of $H_3S$ and $D_3S$ exhibiting highest available values of residual resistance ratio $RRR$ determined for reference temperatures 203.5 K and 0 K. $\Delta T_C/T_C$ is the fractional width of the superconducting transition.

|        | $P$ (GPa) | $T_C$ (K) | $RRR$ | $\Delta T_C/T_C$ |
|--------|-----------|-----------|-------|-------------------|
| $H_3S$ | 141       | 194.4     | 2.261 | 0.0232            |
| $H_3S$ | 155       | 195.8     | 2.283 | 0.0324            |
| $D_3S$ | 155       | 146.6     | 1.617 | 0.0478            |
| $D_3S$ | 173       | 158.6     | 1.658 | 0.0717            |

The measurements of $RRR$ provide an ideal basis for comparing observed $T_C$ between $H_3S$ and $D_3S$. Figure 1 shows that there are two measurements each for $H_3S$ and $D_3S$ that have high values of $RRR$ and are near their respective $P \approx P_m$. The data for these four measurements are given in table 2.

A linear regression fit to these data of the function,

$$T_C = t_1 + t_2\, RRR ,\qquad (2)$$

yields the results $t_1 = 42 \pm 15$ K, $t_2 = 67 \pm 7$ K, coefficient of determination $R^2 = 0.98$, and $\pm 5$ K fitting error. This fit is used to determine the right ordinate scale for $RRR$ in figure 1, which ranges from 0.311 to 2.623 and aligns with the 60–220 K span of the left ordinate for $T_C$. This display of the data points in figure 1 facilitates judging the similarities in the pressure dependences of $T_C$ and $RRR$, noting that the correlations between $RRR$ and $T_C$ found in figure 3 extend to correlations with $P$.

The two middle rows of table 2 contain measurements for both isotopes at the same $P = 155$ GPa, which enables extrapolating the relationship between the values of $T_C$ for the two isotopes to theoretically equal values of $RRR$, i.e., extrapolating to equality in sample quality. This is accomplished by scaling the transition temperatures, $T_C^{H_3S}$ and $T_C^{D_3S}$, of $H_3S$ and $D_3S$, respectively:

$$T_C^{D_3S} = S\, T_C^{H_3S} . \qquad (3a)$$

The scaling parameter $S$ is determined from equation (2) and the residual resistance ratios, $RRR_{H_3S}$ and $RRR_{D_3S}$ for $H_3S$ and $D_3S$, respectively, at matching values of $P$ according to,

$$S = (RRR_{D_3S} + t_1/t_2) / (RRR_{H_3S} + t_1/t_2) . \qquad (3b)$$

With the parameters $t_1$ and $t_2$ from the fit of equation (2) and data for $RRR$ at $P = 155$ GPa in table 2, one obtains the result $S = 0.770 \pm 0.108$.

A measurement of $RRR$ is also available for each isotope in the region $P > P_m$ at comparable values of $P$ (see figure 1); the data are $P = 195$ GPa, $T_C = 181.3$ K, and $RRR = 1.938$ for $H_3S$ and $P = 190$ GPa, $T_C = 149.6$ K, and $RRR = 1.339$ for $D_3S$. For these two points, the formula of equation (3b) evaluates to $S = 0.765$. Hence, the implication is that $S$ is a constant in the region $P \gtrsim P_m$, where $T_C$ decreases with increasing $P$.

**Isotope effect exponent**

Having determined the scaling parameter $S$, the mass isotope effect exponent $\alpha$ may be calculated under the criterion of holding $RRR$ as theoretically constant, i.e., treating $H_3S$ and $D_3S$ as with equivalent order (or disorder) for measurements of $T_C$ at equal applied pressures. From the definition of the mass isotope exponent in $T_C$, the expression applied to the data is,

$$\alpha = \ln(S\, T_C^{H_3S}/T_C^{D_3S}) / \ln(2) . \qquad (4)$$

For the data at $P = 155$ GPa in table 2, equation (4) evaluates to $\alpha = 0.043 \pm 0.140$. For the data at $P = 190\text{–}195$ GPa cited above, one obtains $\alpha = -0.11 \pm 0.14$.

Extension of the calculation to obtain pressure dependence with an assumed fixed $S = 0.77$ is shown in figure 4 according to two methods of treating the $T_C$ data in figure 1. In figure 4(a), the expression of equation (4) is evaluated from the $T_C(P)$ data of a given isotope and the interpolated $T_C$ data of the other isotope. The results cover the range 138 GPa $\leq P \leq$ 226 GPa where the data for $P$ are overlapping. The pronounced increase in apparent $\alpha$ occurring at low pressure comes about because $T_C^{D_3S}$ falls off more rapidly with reduced $P$ than does $T_C^{H_3S}$.

The correlated decreases in $RRR$ and $T_C$ shown in figures 1 and 3 for $H_3S$ at $P < P_m$ suggests that extrinsic disorder depresses the observed $T_C$ in such samples. Although $RRR$ data for $P < P_m$ are presently available only for $H_3S$, the similarities shown in $T_C$ for $P < P_m$ suggests that degradation of sample quality for both $H_3S$ and $D_3S$ underlies the anomalous increase in $\alpha$ at low pressure. In view of the isotope shift in $P_m$, an approach to testing this hypothesis for the region $P < P_m$ is to evaluate equation (4) with $T_C^{H_3S}$ at a given $P$ and interpolated $T_C^{D_3S}$ at $P+P_{\text{offset}}$, and vice versa with $T_C^{D_3S}$ at a given $P$ and interpolated $T_C^{H_3S}$ at $P - P_{\text{offset}}$. Results obtained for $P_{\text{offset}} = 19.8$ GPa based on table 1 are shown in figure 4(b), where the green and blue circles denote results for $H_3S$ and $D_3S$ at given $P$, respectively. This treatment essentially eliminates the low-$P$ upturn in $\alpha$, giving on average $\alpha = 0.13 \pm 0.11$ for $P \lesssim 160$ GPa, a result consistent with the hypothesis and a constant $S$, albeit presumed to be 0.77. A theoretical $\alpha = 0.25$ calculated for a low-$T_C$ phase [7] may be compared to this result.

In the regions corresponding to $P \gtrsim P_m$ for both $H_3S$ and $D_3S$, the $T_C$ data are more weakly dependent on $P$ (factor ~10 smaller slopes, table 1) such that the results for $\alpha$ are presumably less influenced by variations in disorder with $P$. This region corresponds to the data for $P > 160$ GPa in

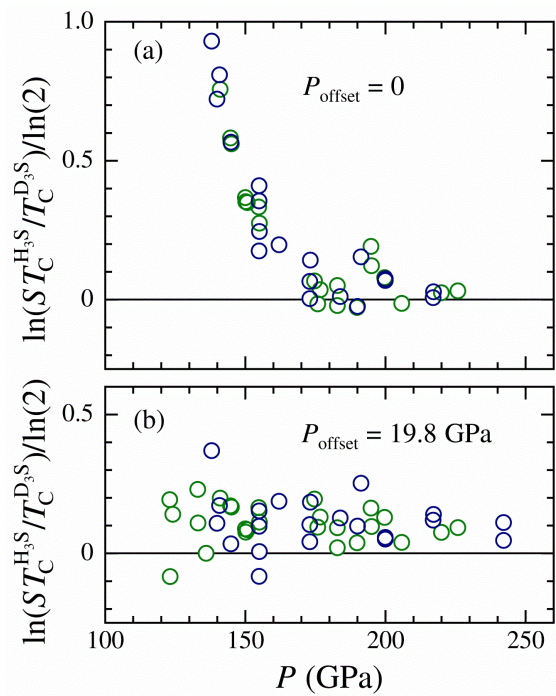

**Figure 4.** Mass isotope effect formula in equation (4) versus applied pressure $P$; frames (a) and (b) correspond to $P_{\text{offset}} = 0$ and 19.8 GPa, respectively. Green circles are from $H_3S$ data and interpolated $D_3S$ data; blue circles are from $D_3S$ data and interpolated $H_3S$ data.

figure 4(a); considering that such results are derived directly from $T_C$ data, one obtains on average $\alpha = 0.048 \pm 0.068$, where the uncertainty is the statistical standard deviation.

## Conclusion

An experimentally accurate value of $\alpha = 0.043 \pm 0.140$ is derived from measurements of transition temperature $T_C$ and residual resistance ratios $RRR$ for $H_3S$ and $D_3S$ at applied pressure $P = 155$ GPa, which is in the region yielding maximum $T_C$. The value of $\alpha$ as defined in equation (4) depends upon the scaling parameter $S$, which is in turn determined from $RRR$ data. The obvious inverse correlation between $RRR$ and the fractional transition width $\Delta T_C/T_C$, as shown in table 2, supports identifying the former as an accurate measure of relative sample order. The aforecited theoretical predictions [4-6], about twice the uppermost limit $\alpha \leq 0.183$ obtained from the data, are in conflict with an experimental $\alpha$ for which the uncertainty also implies consistency with zero. It is evident that materials-based isotopic effects play critically important roles in determining $T_C$, particularly for $D_3S$, which have heretofore been neglected. The isotope effect in $P_m$ and the associated depressed $T_C$ and $RRR$ for $P < P_m$ appear to reflect differing R3m-Im$\bar{3}$m transition pressures [5]. There is also an isotope effect in room temperature annealing, suggesting that a comparatively higher kinetic barrier allows for higher residual disorder in $D_3S$, relative to $H_3S$. As deduced from $RRR$ data for $H_3S$, the apparent anomalous increase in $\alpha$ at low pressures is affected by degraded sample quality at low pressures. The relatively lower values of $RRR$ measured for $H_3S$ and $D_3S$ at $P = 190$–195 GPa suggests decreased $T_C$ at higher pressures correlates with reduced sample quality (and a degraded superconducting state) as well. Other observations corroborating issues with materials quality are 7–8 K reproducibilities in $T_C$ measurements, differences between resistance and magnetic measurements of $T_C$, depressed zero-resistance temperatures, and the absence of field-cooled flux exclusion below $T_C$.

## Acknowledgements

The authors are grateful for support from the College of William and Mary, the New Jersey Institute of Technology, and the University of Notre Dame.